\documentclass[prl,superscriptaddress,twocolumn,preprintnumbers,amsmath,aps]{revtex4}
\usepackage{amsmath,amssymb,color}
\usepackage{epsfig,graphicx,hyperref}
\usepackage{dcolumn}
\usepackage{bm}

\begin{document}
\title{Fluctuation-dissipation relations and \\ critical quenches in the transverse field Ising chain}
\author{Laura Foini} 
\affiliation{SISSA -- International School for Advanced
  Studies and INFN, via Bonomea 265, 34136 Trieste, Italia}
\affiliation{Laboratoire de Physique Th\'eorique,
  Ecole Normale Sup\'erieure de Paris,  UMR 8549, 
24 rue Lhomond 75234 Paris
  Cedex 05, France} 
\author{Leticia F. Cugliandolo}
\affiliation{Universit\'e Pierre et Marie Curie -- Paris VI, Laboratoire de Physique Th\'eorique
et Hautes Energies, UMR
  7589, Tour 15 5\`eme \'etage, 4 Place Jussieu, 75252 Paris Cedex 05, France} 
\author{Andrea  Gambassi} 
\affiliation{SISSA -- International School for Advanced
  Studies and INFN, via Bonomea 265, 34136 Trieste, Italia}
\affiliation{Universit\'e Pierre et Marie Curie -- Paris VI, Laboratoire de Physique Th\'eorique
et Hautes Energies, UMR
  7589, Tour 15 5\`eme \'etage, 4 Place Jussieu, 75252 Paris Cedex 05, France} 

\date{\today}

\newcommand{\rme}{{\rm e}}
\newcommand{\rmd}{{\rm d}}
\newcommand{\eff}{{\rm eff}}
\newcommand{\ie}{{\it i.e.}}
\newcommand{\reff}[1]{(\ref{#1})}
 \let\Y=\Upsilon

\begin{abstract}
Dynamic correlation and response functions of classical and quantum systems in thermal equilibrium are connected by fluctuation-dissipation theorems, 
which allow an alternative definition of their (unique) temperature.
Motivated by this fundamental property, we revisit the issue of thermalization of closed many-body quantum systems long after a sudden quench, focussing on the non-equilibrium dynamics of the Ising chain in a critical transverse field. 
We show the emergence of distinct observable-dependent effective temperatures, which rule out Gibbs thermalization in a strict sense but might still have a thermodynamic meaning.
\end{abstract}

\maketitle
\setlength{\textfloatsep}{10pt} 
\setlength{\intextsep}{10pt}

\noindent\emph{Introduction.} The development of experimental techniques which prevent
dissipation in quantum many-body systems has triggered increasing interest in
the non-equilibrium dynamics of such closed systems.
The unitary non-equilibrium dynamics of a system initially prepared in a state which
is not an eigenstate of its Hamiltonian is called a  \emph{quantum quench}. 
Basic questions as to
whether a stationary state is reached and how this can be
characterized naturally arise. These questions have been addressed in a
number of simple models, including the one-dimensional systems reviewed
in Refs.~\cite{Karevski,reviews-chains}. Early studies led to the following picture:
Non-integrable systems should eventually reach a thermal stationary state characterized 
by a Gibbs distribution with a single temperature. 
Integrable systems, instead, are not expected to thermalize 
but their asymptotic stationary state should nonetheless
be described by a so-called generalized Gibbs
ensemble (GGE) with one effective temperature for each conserved
quantity~\cite{Rigol,GGE_other_models,GGE_Ising,Calabrese11}. 
Interestingly enough, depending on the specific quantity and the system's parameters
a Gibbs ensemble turns out to capture anyhow some relevant features 
of the non-equilibrium dynamics of integrable systems~\cite{Rossini10}. In 
particular, observables that are non-local in the quasi-particles display numerically 
the same relaxation scales as in equilibrium with a suitable effective temperature,
at least for small quenches~\cite{Rossini10,Calabrese11}. Local quantities
instead do not, with possible exceptions for quenches at criticality.

Our purpose is to revisit the debated issue of thermalization in closed quantum systems 
with tools developed for the study of \emph{classical
and quantum dissipative glassy systems}.  The analysis of thermalization in closed 
quantum systems focused so far on the property that   
expectation values of quantities --- such as (i) the conserved energy and
two-point correlation functions depending on either (ii) one 
or (iii) two times --- should behave, at long times, as the corresponding averages
calculated on suitable statistical ensembles. However, an equally important 
property of thermal equilibrium states of 
both classical and quantum systems is the validity of model-independent 
fluctuation-dissipation theorems (FDTs)~\cite{Kubo}, which relate 
\emph{linear response and correlation functions} independently of their 
functional form. Focusing on two-time quantities, we investigate 
thermalization issues from this perspective. 

Before getting into the technical details, let us explain why 
fluctuation-dissipation relations (FDRs) should be more 
relevant to thermalization issues 
than  the precise functional decay of observables. Take phase separating systems
as an example.
The expectation values of one-time quantities -- such as the energy density (i) or observables of the type (ii)  --
reach equilibrium values, suggesting the equilibration of the 
sample at the bath temperature. This is, however, incorrect as 
proven by the fact that  observables of type (iii), such as delayed density 
correlations and linear responses, decay in more than one dynamic scale, 
and in the slowest they do 
algebraically, as opposed to the typical exponential
equilibrium relaxation. 
In spite of this, one can still define a {\it bona fide} effective
temperature from FDRs that link correlations to their associated linear 
response~\cite{Cugliandolo97} as long as one distinguishes different pairs
of  observables and the time-scales in which they evolve. 
Indeed, some observables basically ignore others --- they do not interact --- 
equilibrate quickly with the environment and are characterized by its 
temperature. This is the case of particle velocities. Positions, instead, 
do not equilibrate with the bath but acquire the same effective temperature, \ie, 
they partially equilibrate, in their own common regime of relaxation.  This notion 
applies to many other systems with slow dynamics and 
constitutes  the basis of, {\it e.g.}, a consistent thermodynamic picture of non-equilibrium glassy
dynamics as realized in mean-field theory~\cite{Teff-reviews}. A quantum finite-dimensional  
example with effective temperatures are electronic tight-binding 
dissipative one-dimensional models driven by external 
fields~\cite{Caso}.

We explore up to which extent these ideas carry 
over to the non-equilibrium dynamics 
of a highly excited closed quantum system. 
For concreteness, we focus on critical quenches in which the
system's parameters for times $t>0$ are tuned to be at an equilibrium (quantum) critical point. 
The detailed analysis of the effective
temperature based on FDRs in classical thermal quenches~\cite{Teff-reviews,Teff-critical}
demonstrated that such a parameter has a thermodynamic meaning for very late
epochs only~\cite{Godreche}, when one of the involved times is much longer than the
other (see, however, Ref.~\cite{Andrea}).

\noindent\emph{The model.} To illustrate our approach, we focus on the simplest quantum integrable 
interacting model, the transverse field Ising chain~\cite{reviews-chains}
\begin{equation}
{\cal H}_\Gamma = - J\; \mbox{$\sum_{i=1}^L$} \;\left( \;\sigma_i^x \sigma_{i+1}^x + \Gamma ~\sigma_i^z\;\right) \;
\label{eq:model}
\end{equation}
with periodic boundary conditions and even $L$. 
The Pauli matrices $\sigma_i^{x,z}$ satisfy the 
SU(2) algebra on the same site $i$ and commute on different sites.  
In what follows  we set $J,\hbar,k_B =1$
and we measure time in units
of $\hbar/J$ and the temperature $T$ in units of $J/k_B$.
The model is exactly solved by a Jordan-Wigner transformation 
to free fermions followed by a Bogoliubov rotation in momentum space~\cite{mccoy}. The
energy of the elementary fermionic excitations with momentum $k$ is
$\epsilon_k(\Gamma)$ $=$  $2 [\Gamma^2-2 \Gamma \cos k +1]^{1/2}$.  For $T=0$ and 
$L\to\infty$ a quantum critical
point  at $\Gamma=1$ separates a paramagnetic 
phase (PM, $\Gamma>1$) with  
$\langle\sigma_i^x\rangle = 0$,
from a ferromagnetic phase (FM, $\Gamma < 1$) with spontaneous symmetry breaking
$\langle\sigma_i^x\rangle \neq 0$ and long-range order along the $x$ direction.
$\langle \sigma_i^z\rangle \neq 0$ for all $\Gamma > 0$.

The system is prepared at $t=0$ in the ground state 
$|\psi_0\rangle$ of ${\cal H}_{\Gamma_0}$, while it subsequently evolves with 
${\cal H}_{\Gamma=1}$, \ie, at the critical point.
The quench from $\Gamma_0$ to $\Gamma$ injects an extensive amount of 
energy  into the system which is henceforth conserved. 
After a transient (studied in Refs.~\cite{Igloi00,Igloi11} for the chain with free boundaries)
the system reaches an asymptotic stationary regime.
A crucial quantity in the description of the dynamics  
is the difference $\Delta_k$ between the Bogoliubov angles 
diagonalizing ${\cal H}_\Gamma$  and ${\cal H}_{\Gamma_0}$:
\begin{equation}
\cos \Delta_k(\Gamma ,\Gamma_0) = 
\frac{4 ~[\Gamma \Gamma_0 - (\Gamma+\Gamma_0) \cos k + 1]}{\epsilon_k(\Gamma) \epsilon_k(\Gamma_0)} 
\; . 
\label{eq:Delta_k}
\end{equation}
$\Delta_k$ encodes the dependence on the initial state and fixes the non-thermal statistics of the excitations created 
at $t=0$. 

\begin{figure}[h]
\centering
 \includegraphics[scale=0.23]{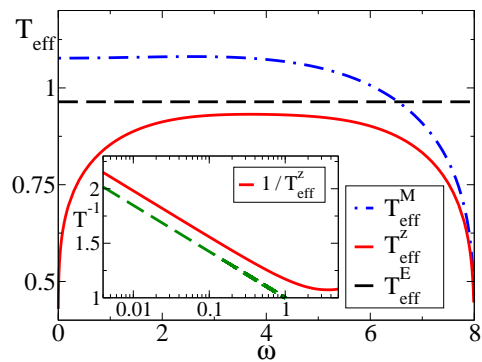} 
\caption{(Color online.)  Comparison between the effective temperature 
defined via the frequency-domain FDR 
applied to $\sigma_i^z$ 
(red solid line) and  $M$ (blue dash-dotted line), and 
 the energy definition (black dashed horizontal line) for
$\Gamma_0=0.3$. The inset highlights the
logarithmic divergence of $T_{\rm eff}^z(\omega)$ for $\omega\to 0$.}
\label{fig:beffz-omega}
\end{figure}

\begin{figure}[h]
\centering
 \includegraphics[scale=0.21]{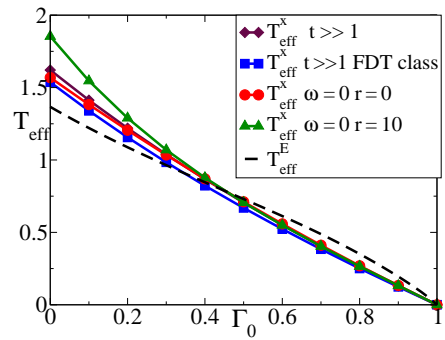} 
\caption{(Color online.) 
$\Gamma_0$ dependence of 
the order parameter effective temperature $T^x_{\rm eff}$
 compared to $T^E_{\rm eff}$ defined from the energy [see Eq.~\reff{Teff_energy}] (dashed line). 
The solid lines, from bottom to top, indicate the values determined 
on the basis of the classical limit of the FDR in the time domain, 
of the limit $\omega\to 0$ of the frequency-domain FDR,
of Eq.~(\ref{Teff_x_time}), and of the limit $\omega\to 0$ of the frequency-domain FDR
but for spins separated by a distance $r=10$. The dependence on $\Gamma_0$
can be read from the one of $\tau$ and Eq.~(\ref{Teff_x_time}).
}
\label{fig:Teff-Gamma0}
\end{figure}

A criterion that has been used to define an effective 
temperature $T^{E}_{\rm eff}(\Gamma,\Gamma_0)$ is to require that  
the energy after the quench --- quantity (i) in the Introduction --- equals the average
over a thermal equilibrium ensemble~\cite{reviews-chains,Rossini10}.  
For model \reff{eq:model} this implies~\cite{Rossini10}
\begin{equation}
0= \!\int_0^{\pi}\! \frac{\rmd k}{\pi} \epsilon_k(\Gamma) \! \left[ \cos \Delta_k(\Gamma,\Gamma_0) 
- \tanh\frac{\epsilon_k(\Gamma)}{2 T_{\rm eff}^E(\Gamma,\Gamma_0)} \right]
\label{Teff_energy}
\end{equation}
which results in the $T_{\rm eff}^E$ shown
in Figs.~\ref{fig:beffz-omega} and~\ref{fig:Teff-Gamma0} with a dashed
black line.  Requiring, instead, the integrand in Eq.~\reff{Teff_energy} to vanish defines the mode-dependent 
$T_{\rm eff}^{\epsilon_k}(\Gamma,\Gamma_0)$ of the 
GGE~\cite{Calabrese11,GGE_Ising}. 

\noindent\emph{FDT.} We focus on the symmetrized and anti-symmetrized two-time correlations
of two  
operators $A$ and $B$ in the Heisenberg representation,
$A_H(t) = \rme^{i{\cal H}_\Gamma t} A \rme^{-i {\cal H}_\Gamma t}$, 
\begin{equation}
\label{Cpm}
C^{AB}_{\pm}(t+t_0,t_0) = \langle \psi_0 | [A_H(t+t_0), B_H(t_0)]_\pm |\psi_0\rangle, 
\end{equation}
where $[X,Y]_\pm \equiv (XY \pm YX)/2$.
More precisely, we consider connected correlations
that we still denote by $C^{AB}_{\pm}$.  While $C^{AB}_+$ 
approaches the classical correlation function for $\hbar\to 0$,
$C^{AB}_-$ is related to the
linear response function $R^{AB}$ through the 
Kubo formula
$R^{AB}(t_1,t_0)=\delta \langle A_{H-h(t) B} (t_1)\rangle_H/\delta h(t_0)\Big|_{h=0} \!\! =  (2i/\hbar)
C_-^{AB}(t_1,t_0) \theta(t_1-t_0)$~\cite{Kubo} which is valid in and out of equilibrium
[$\theta(t\!<\!0)=0$ and $\theta(t\!>\!0)=1$].
The asymptotic stationary regime is formally defined by the limit
$t_0 \to \infty$. When it is physically established depends on the observable and 
$\Gamma_0$.
Natural choices for $A$ and $B$ are the order parameter
$\sigma_i^x$ and the transverse magnetization $\sigma_i^z$. The
$\sigma_i^x$ autocorrelation functions,  
$C_\pm^x(t)$, are non-local with respect to the
quasi-particles while those of $\sigma_i^z$, 
$C_\pm^z(t)$, are local
in the same variables~\cite{reviews-chains}.  

In Gibbs equilibrium at inverse temperature $\beta$ 
the FDT connects $C^{AB}_+$ and $R^{AB}$, via the model-independent 
time-domain and frequency-domain relations~\cite{Kubo}
\begin{eqnarray}
&& 
R^{AB}(t) = \frac{i}{\hbar} 
\int_{-\infty}^{\infty} \frac{\rmd\omega}{\pi} \rme^{-i\omega t} \tanh(\beta\hbar\omega/2) 
\, \tilde C^{AB}_+(\omega), 
\quad
\label{eq:FDT-t}
\\
&&
\hbar~\mbox{Im} \tilde R^{AB}(\omega) =
\tanh(\beta\hbar \omega/2) \,
\tilde C^{AB}_+(\omega), 
\label{eq:FDT-omega}
\end{eqnarray}
respectively. We reinstated $\hbar$
to make the classical limit, $R^{AB}(t) = -\beta \rmd C^{AB}(t)/\rmd t
\ \theta(t)$, transparent.  The FDR definition of an effective temperature 
amounts to replacing $\beta$ by $\beta_{\rm eff}(\omega)$ in Eqs.~(\ref{eq:FDT-t}) 
and (\ref{eq:FDT-omega}) [with a $t_0$-dependence  in non-stationary
(glassy) cases], \ie,
\begin{equation}
\hbar~\mbox{Im} \tilde R^{AB}(\omega) =
\tanh[\beta_{\rm eff}^{AB}(\omega) \hbar \omega/2] \
\tilde C^{AB}_+(\omega)
\; . 
\label{eq:Teff-def-freq}
\end{equation} 
This is the definition that we shall repeatedly use below.

Before presenting our results 
let us summarize what is known about $C_\pm^{z,x}$ for $\Gamma=1$.
In equilibrium ($\Gamma_0=\Gamma$)  
$\langle \psi_0 | \sigma_i^z(t+t_0)\sigma_i^z(t_0)|\psi_0\rangle$
decays algebraically as $|t|^{-3/2}$ at $T=0$ and as
$|t|^{-1}$ at finite $T$~\cite{Rossini10}.   Out of equilibrium the
decay of $C_+^z$ is $|t|^{-2}$ ~\cite{Igloi00}.  For the special case of a fully polarized
initial condition $(\Gamma_0= \infty$) 
$C_-^z$ follows the same $|t|^{-2}$  decay, as can be inferred from
the results in Ref.~\cite{Karevski}.
Instead, a generic
exponential relaxation of $\langle \psi_0 |
\sigma_i^x(t_0)\sigma_j^x(t_0)|\psi_0\rangle$
was argued in~\cite{Calabrese06} using semi-classical methods and
later shown to hold exactly~\cite{Calabrese11}.  This is
in contrast to the power-law decay of the $T=0$ equilibrium
order-parameter spatio-temporal correlations.  
As far as we know, $C^x_-$ has not been analyzed so far. 
Here we complete this picture by calculating
$C^x_\pm$ and $C^z_\pm$ for generic $\Gamma_0$. 
We also study $C_\pm^M$, 
where $A,B=M=\sum_{i=1}^L \sigma_i^z/L$. 

\noindent{\it The transverse local magnetization.}
For $\Gamma_0\neq\Gamma=1$ we found that 
$C_+^z$ and $R^z$ decay as a sum of
power laws of $t$ and $t+2t_0$.
Thus, no characteristic time can be identified and, in addition, 
one cannot compare these functions to the
thermal ones in order to define an effective temperature, 
as done in Refs.~\cite{Rossini10,Igloi11,Calabrese11} for 
other observables [type (iii) of the Introduction]. 
Taking $t_0\to\infty$ one finds the stationary relaxation
\begin{eqnarray}
\label{CZ}
&&
C^z_+(t) = -(8\pi t^2)^{-1} \cos 8t + {\cal O}(t^{-3})
\; , 
\\
&&
R^z(t) \, = 
\ (4\pi t^2)^{-1} \left[  \Y^{-1} - \sin 8t  \right]
+{\cal O}(t^{-3}) \; ,
\label{RZ}
\end{eqnarray}
with $\Y=[(1+\Gamma_0)/(1-\Gamma_0)]^2$.
The complementary analysis  in the frequency domain allows us to 
define a frequency-dependent $T_{\eff}^z$ via the FDR in Eq.~(\ref{eq:Teff-def-freq}).
The function $T_{\eff}^z(\omega)$ is shown in
Fig.~\ref{fig:beffz-omega}~for $\Gamma_0=0.3$ as a (red) solid
line and it has to be compared with the
constant value obtained from 
Eq.~(\ref{Teff_energy}), as in Ref.~\cite{Rossini10}, shown as a black dashed line. 
The asymptotic regime corresponds to the limit $\omega
\to 0$, zoomed in the inset, 
in which $\beta_{\eff}^z = 1/T_{\eff}^z$ diverges logarithmically with the
law $ -1/2 (1-1/\Y)/[1 + (\Y-2) \arctan(\sqrt{\Y-1})/\sqrt{\Y-1}] \ \ln \omega$  (green dashed line).  
We conclude that, although $\langle \sigma^z_i\rangle$ takes a thermal value~\cite{Rossini10}, the dynamics of
$\sigma^z_i$ is not compatible with an equilibrium thermal behavior.
For other values of $\Gamma_0$, 
$T_{\eff}^z$  still vanishes at $\omega=0$ and $\omega=\omega_{\rm max}\equiv 8$ but it is not concave
for $\Gamma_0\gtrsim 0.35$~\cite{laura-prep}.
For increasingly
narrower quenches with 
$\Gamma_0\to 1$,
$\beta_{\eff}^z\to\infty$ uniformly over all frequencies.

\noindent{\it The transverse global magnetization.}
The long-time stationary decay of the global magnetization 
correlations are even slower than the one of $C_{\pm}^z(t)$ \cite{notaM}: 
\begin{eqnarray}
\label{CM}
&&
C_{+}^M(t) = (8\sqrt{\pi}t^{3/2})^{-1}  \sin(8t - \pi/4) + {\cal O}(t^{-5/2}), 
\label{CM}\\
&&
R^M(t) = - (4\sqrt{\pi}t^{3/2})^{-1} \cos(8t - \pi/4)
+ {\cal O}(t^{-5/2}).\qquad
\label{RM}
\end{eqnarray}
The leading-order decay $t^{-3/2}$ 
is the same as in equilibrium at finite $T$. However, while the prefactor 
depends upon $T$ in equilibrium \cite{Niemeijer}, out of equilibrium 
the dependence on $\Gamma_0$ appears only at the next-to-leading order, \ie,
the long-$t$ limit of $C_{\pm}^M(t) $ does not retain memory of the
initial condition. 

The  (blue) dash-dotted line in Fig.~\ref{fig:beffz-omega} is $T_{\rm eff}^M(\omega)$ 
as obtained  from the FDR \reff{eq:Teff-def-freq}, applied to  
$C_\pm^M$. $T_{\eff}^M(\omega)$ approaches
$2/\sqrt{\Y}$ at low frequencies, it vanishes for $\omega=\omega_{\rm max}$ and becomes 
non-monotonic for $\Gamma_0\gtrsim 0.27$, developing a shallow local maximum.
Naively, one may expect to recover this value by 
treating the time-domain FDR in the long-$t$ limit as follows.
Replacing $\beta$  by a constant effective value 
$\beta^*_{\rm  eff}$ in the rhs of Eq.~\reff{eq:FDT-t}, the integral can be written as series of odd time 
derivatives of $C_+^M(t)$. Inserting Eqs.~\reff{CM} and~\reff{RM} in the rhs and lhs of this 
expression, respectively, yields $1 = \tanh(4 \beta^*_{\rm eff})$  for $t\to\infty$ and therefore 
$T^*_{\rm eff}= 0$. The fact that $T_{\eff}^M(\omega\!\to\!0) \neq T^*_{\rm eff}$ indicates that 
$\beta^M_{\rm eff}(\omega)$ cannot be approximated by an average constant in the integral. 
Indeed, since only the derivatives of the oscillating factor in~\reff{CM} 
contribute to the leading order of Eq.~\reff{eq:FDT-t}, $T^*_{\rm eff}$ is the one 
of the oscillatory frequency, which coincides with the threshold value $\omega_{\rm max}$ 
and, for $\omega\to\omega_{\rm max}$,  $\beta_{\rm eff}^M$ diverges as $\beta_{\rm eff}^M(\omega) \simeq - 
\ln(\omega_{\rm max} - \omega) / 4$. 
Such threshold results from the maximum of the dispersion relation and the quadratic dependence 
of $M$ on the fermionic excitations. It is therefore unclear whether $T_{\eff}^M(\omega\!=\!0)$ can be recovered from the FDR in the time domain.

Interestingly enough, it turns out \cite{laura-prep} that each frequency $\omega$ selects a 
mode $k$ such that $2\epsilon_k = \omega$ and $T_{\eff}^{M}(\omega)$ 
defined here from the FDR \reff{eq:Teff-def-freq} coincides with the 
temperature $T_{\eff}^{\epsilon_k} $ of the GGE. 
%
%
%
%

\begin{figure}[h]
\centering
 \includegraphics[scale=0.23]{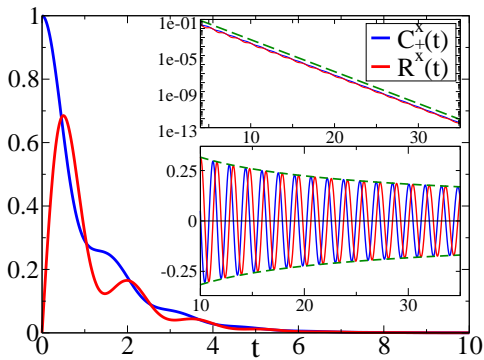}
\caption{(Color online.) Decay of the order-parameter
  correlation $C_+^x$ (blue line) and the linear response
  $R^x$ (red line) for $\Gamma_0=0.3$. Upper inset: zoom into the long-$t$ 
  decay that demonstrates the exponential relaxation with the characteristic 
  time $\tau$ defined in the text (dashed green line). Lower inset:
  $(\rme^{t/\tau}C_+^x/A_C -1)/a_C$ and $(\rme^{t/\tau}R^x/A_R-1)/a_r$ 
  vs. $t$; the green dashed line is the $t^{-1/2}$
  envelope of the damped oscillations, in agreement with Eqs.~\reff{eq:Cx-fit} and~\reff{eq:Rx-fit}.
}
\label{fig:RCz}
\end{figure}

\noindent{\it The order parameter.} %
Equations~\reff{CZ}, \reff{RZ}, \reff{CM} and \reff{RM} are invariant under 
$\Gamma_0\mapsto\Gamma_0^{-1}$ because
$\cos\Delta_k(1,\Gamma_0)=\cos \Delta_k(1,\Gamma^{-1}_0)$ 
[see Eq.~\reff{eq:Delta_k}]
is the sole quantity bringing about the dependence on $\Gamma_0$ in the stationary limit 
$t_0\to\infty$ of $C^{M,z}_\pm$ \cite{laura-prep}. In the same limit 
and for $\Gamma=1$ we find numerically that this
invariance also  holds for $C_{\pm}^x$. Henceforth we
restrict to quenches originating from the FM
phase.  
We computed $C_{\pm}^x$  for a chain with $L=10^3$ and  $t_0=10$ with the methods employed in 
Refs.~\cite{mccoy,Rossini10,Igloi00}.
Our numerical results are fitted very accurately by
\begin{eqnarray}
&&
C^x_+(t) \simeq \rme^{-\frac{t}{\tau}} A_C [ 1 + a_C ~t^{-\frac{1}{2}} \ \sin (4t+\phi) ], 
\label{eq:Cx-fit}
\\
&& 
R^x(t) \simeq \rme^{-\frac{t}{\tau}} A_R [1 - a_R~ t^{-\frac{1}{2}} \ \cos (4t+\phi) ],
\label{eq:Rx-fit}
\end{eqnarray}
with (numerically) the same rate $\tau^{-1}$ as the one 
$\tau^{-1} = - \int_0^\pi (\rmd k/\pi) [\rmd\epsilon_k(\Gamma)/\rmd k] \ln\cos\Delta_k$
analytically proved \cite{Calabrese11} to characterize the exponential long-time decay of $ \langle \psi_0 |\sigma^x_i(t)\sigma^x_j(t) | \psi_0 \rangle$.
The expression for $\tau$ finds further support from the fact that with the substitution
$\cos\Delta_k \to \tanh(\beta\epsilon_k/2)$ one recovers the
equilibrium $\tau_{\rm eq}$~\cite{Deift,nota} in this as well as other
statistical averages.
Although several fitting parameters are involved in Eqs.~\reff{eq:Cx-fit} and \reff{eq:Rx-fit},
we tested these expressions in various instances and they turned out to be always 
remarkably accurate already for $t\gtrsim 5$~\cite{laura-prep}.
In Fig.~\ref{fig:RCz} we show $C^x_+$ and $R^x$
for $\Gamma_0=0.3$, together with a zoom into the
long-time decay and its comparison with the leading exponential decay
in the upper inset. The lower inset 
confirms the high quality of the
fit of the correction terms to the forms given in
Eqs.~\reff{eq:Cx-fit} and \reff{eq:Rx-fit} which are actually indistinguishable from the 
data.  The non-equilibrium coherence time 
$\tau = \pi \sqrt{\Y-1}/[4  \arctan(\sqrt{\Y-1})]$
decreases upon increasing  $|1-\Gamma_0|$, \ie, the energy injected into
the system and $\tau \sim |1-\Gamma_0|^{-1}$ for $\Gamma_0 \to 1$.
While the parameters $A_{R,C}$ depend on $\Gamma_0$, 
their ratio $A_C/A_R=1.210(5)$ does not
within our numerical accuracy. 
More details on the fitting parameters of the oscillating (lattice) 
correction will be presented in Ref.~\cite{laura-prep}.

%
%
%
%

The effective temperatures determined in the frequency and in the time domain 
are equivalent in this case.
As discussed above, replacing
$\beta$ by a constant 
$\beta_{\eff}^x$, turns the rhs of the FDT \reff{eq:FDT-t} in the time domain  into a  
series of time derivatives of $C^x_+(t)$ which yields
\begin{equation}
\label{Teff_x_time}
\hbar A_R/(2A_C) =  \tan(\hbar\beta_{\eff}^x/2 \tau)
\end{equation}
for $t\to\infty$, \ie, neglecting the oscillatory 
corrections in Eqs.~\reff{eq:Cx-fit} and \reff{eq:Rx-fit}.
Alternatively, Eq.~\reff{eq:Teff-def-freq} yields
$\beta_{\eff}^x(\omega=0) = \int_0^{\infty}\!\rmd t\, t R(t) / \int_0^{\infty}\!\rmd t\,C(t) $
 for $\omega \to 0$
which numerically coincides with the constant value in Eq.~\reff{Teff_x_time}.
For $\hbar\beta_{\rm eff}^x/2\tau \ll 1$ one recovers 
the classical limit
$\beta^x_{\eff} \simeq -R^x(t)/[\rmd C^x(t)/\rmd t]
\simeq \tau A_R/A_C$. 
All three determinations of $T_{\rm eff}^x$ are shown in
Fig.~\ref{fig:Teff-Gamma0} as functions of
$\Gamma_0$ and they are
compared to 
$T^E_{\rm eff}$ (dashed line) from Eq.~(\ref{Teff_energy})~\cite{Rossini10}.
%
%
%
%

We completed our analysis by studying space-dependent correlations 
and we found that they yield analogous results, as shown in 
Fig.~\ref{fig:Teff-Gamma0}
for the $x$-component of two spins at distance $r=10$. (Note that differently from 
case $r=0$ shown in Fig.~\ref{fig:RCz}, 
correlations   with $r\neq 0$
display a light-cone effect due to the finite speed 
of the quasi-particles~\cite{Calabrese06,Calabrese11,Igloi11,laura-prep}.)

\noindent\emph{Conclusions.}  
Independently of the functional form 
of the correlations involved, the FDRs allow us to define various effective temperatures. 
We calculated the (self) FDR for three observables that are 
local ($\sigma_i^{x,z}$) or non-local ($M$) in space and local 
($\sigma_i^z$, $M$) or non-local ($\sigma_i^x$) in the quasi-particles.
$\sigma_i^z$ is not compatible with Gibbs thermal equilibrium at any effective temperature. 
The frequency-domain FDR for $M$ yields a finite $T^M_{\rm eff}(\Gamma_0)$ 
in the limit $\omega \to 0$. Frequency and time-domain 
determinations of $T_{\rm eff}^x(\Gamma_0)$ are equivalent. 
$T_{\rm eff}^M$ and $T_{\rm eff}^x$ have the same qualitative dependence
on $\Gamma_0$ but they differ (also from $T_{\rm eff}^E$). 
This excludes a single temperature effective Gibbs description 
(as the one discussed in Ref.~\cite{Mitra11}) of the full stationary
dynamics  of this model but the question remains as to whether some of 
the temperatures which emerge can be attributed a 
thermodynamic meaning. 

We finally stress that a \emph{bona fide} thermal behavior should be accompanied 
by the validity of suitable FDTs also in the context of quantum quenches.

\noindent\emph{Acknowledgments.} We thank F. Igl\'oi, G. Santoro, G. Semerjian, A. Silva, 
F. Zamponi for discussions. 
A. Gambassi and L. F. Cugliandolo thank  LPTHE and ICTP, respectively, 
for hospitality during the preparation of this work that 
was financially supported by ANR-BLAN-0346 (FAMOUS) and 
CNRS.


\begin{references}

\bibitem{reviews-chains} A. Polkovnikov, 
K. Sengupta, A. Silva, M. Vengalattore, 
{\tt arXiv:1007.5331}. 
J. Dziarmaga, Adv. in Phys. {\bf 59}, 1063 (2010).

\bibitem{Karevski} 
D. Karevski, 
{\tt arXiv:cond-mat/0611327}.

\bibitem{Rigol} 
M. Rigol, 
V. Dunjko, V. Yurovsky, and M. Olshanii,
Phys. Rev. Lett. {\bf 98}, 050405 (2007).
M. Rigol, 
A. Muramatsu, and M. Olshanii, 
Phys. Rev. A {\bf 74}, 053616 (2006).

\bibitem{GGE_other_models}
M. A. Cazalilla, Phys. Rev. Lett. {\bf 97}, 156403 (2006);
A. Iucci and M. A. Cazalilla, Phys. Rev. A {\bf 80}, 063619 (2009).

\bibitem{GGE_Ising}
D. Fioretto and G. Mussardo, 
New J. Phys. {\bf 12}, 055015 (2010).

\bibitem{Calabrese11} 
P. Calabrese, 
F. Essler, and M. Fagotti,
Phys. Rev. Lett. {\bf 106}, 227203 (2011).

\bibitem{Rossini10} D. Rossini, 
A. Silva, G. Mussardo, and G. E. Santoro, 
Phys. Rev. Lett.  {\bf 102}, 127204 (2009);
D. Rossini, 
S. Suzuki, G. Mussardo, G. E. Santoro, and A. Silva,
Phys. Rev. B {\bf 82}, 144302 (2010).

\bibitem{Kubo} 
R. Kubo, 
M. Toda, and N. Hashitume, 
{\it Nonequilibrium
  statistical mechanics}, 2nd ed. (Springer Verlag, 1991).

\bibitem{Cugliandolo97} 
L. F. Cugliandolo, 
J. Kurchan, and L. Peliti,
Phys. Rev. E {\bf 55}, 3898 (1997).

\bibitem{Teff-reviews} 
A. Crisanti and F. Ritort, J. Phys. A {\bf 36}, R181 (2003).  
L. Leuzzi, J. Non-Cryst. Sol. {\bf 355}, 686 (2009).
L. F. Cugliandolo, 
{\tt arXiv:1104.4901};L. F. Cugliandolo and G. Lozano,
  Phys. Rev. Lett. {\bf 80}, 4979 (1998); Phys. Rev. B {\bf 59}, 915 (1999).

\bibitem{Caso} 
A. Caso, 
L. Arrachea, and G. S. Lozano, 
Phys. Rev. B {\bf 83}, 165419 (2011) and refs. therein.

\bibitem{Teff-critical} 
P. Calabrese and A. Gambassi, 
J. Phys. A {\bf 38}, R133 (2005);
F. Corberi, 
E. Lippiello, and M. Zannetti,
J. Stat. Mech. P07002 (2007).


\bibitem{Godreche}
C. Godr\`eche and J.-M. Luck, J. Phys. A {\bf 33}, 1151 (2000).

\bibitem{Andrea} 
P. Calabrese and A. Gambassi, 
J. Stat. Mech. P07013 (2004); 
J. Stat. Mech. Theor. Exp. P01001 (2007). 


\bibitem{mccoy}
  B. M. McCoy, 
  E. Barouch, and D. B. Abraham, 
  Phys. Rev. A {\bf 4}, 2331 (1971).
  
\bibitem{Igloi00} 
F. Igl\'oi and H. Rieger, 
Phys. Rev. Lett. {\bf 85}, 3233 (2000).


\bibitem{Igloi11}
 F. Igl\'oi and H. Rieger,  Phys. Rev. Lett. {\bf 106}, 035701 (2011);
 {\tt arXvi:1106.5248}.

\bibitem{Calabrese06} 
P. Calabrese and J. Cardy, 
Phys. Rev. Lett. {\bf 96}, 136801 (2006); 
J. Stat. Mech. P06008 (2007).

\bibitem{laura-prep} 
L. Foini, 
L. F. Cugliandolo, and A. Gambassi
(unpublished).



\bibitem{notaM} 
$C_+^M(t)$ has an additional constant term, in and out of equilibrium, equal to  $(\sqrt{\Upsilon}+1)^{-2}$ for critical quenches. It is however irrelevant for the study of FDT. 
Note also that the definition of $C_{\pm}^M(t)$ as in Eq.~\reff{Cpm} requires an extra normalization 
factor $L$, necessary for global observables. 

\bibitem{Niemeijer}
T. Niemeijer, 
Physica {\bf 36}, 377 (1967).


\bibitem{Deift}
P.~Deift and X.~Zhou, in {\it Singular limits of dispersive waves}
(Lyon, 1991), p. 183, NATO Adv. Sci. Inst. Ser. B Phys. 320
(Plenum, New York, 1994).

\bibitem{nota}
The value of $\tau_{\rm eq}$ for small $T$
quoted sometimes in the literature
differs from the correct one $\tau_{\rm eq}=4/(\pi T)$ ~\cite{Deift}.
 

\bibitem{Mitra11} 
A. Mitra and T. Giamarchi, 
{\tt arXiv:1105.0124}.
J. Lancaster, T. Giamarchi, and A. Mitra, 
{\tt arXiv:1105.6039}.

 
\end{references}
\end{document}